%% file: ms.tex
\documentstyle[emulateapj,rotate,times]{article}
 
\newcommand{\etal}{et~al.\ }
\newcommand{\PVdblt}{{\rm P}\kern 0.1em{\sc v}~$\lambda\lambda 1117, 1128$}
\newcommand{\CaIIdblt}{{\rm Ca}\kern 0.1em{\sc ii}~$\lambda\lambda 3934, 3969$}
\newcommand{\AlIIIdblt}{{\rm Al}\kern 0.1em{\sc iii}~$\lambda\lambda 1854, 1862$}
\newcommand{\CIVdblt}{{\rm C}\kern 0.1em{\sc iv}~$\lambda\lambda 1548, 1550$}
\newcommand{\MgIIdblt}{{\rm Mg}\kern 0.1em{\sc ii}~$\lambda\lambda 2796, 2803$}
\newcommand{\NVdblt}{{\rm N}\kern 0.1em{\sc v}~$\lambda\lambda 1238, 1242$}  
\newcommand{\SVIdblt}{{\rm S}\kern 0.1em{\sc vi}~$\lambda\lambda 933, 944$} 
\newcommand{\OVIdblt}{{\rm O}\kern 0.1em{\sc vi}~$\lambda\lambda 1031, 1037$} 
\newcommand{\SiIIdblt}{{\rm Si}\kern 0.1em{\sc ii}~$\lambda\lambda 1190, 1193$} 
\newcommand{\SiIVdblt}{{\rm Si}\kern 0.1em{\sc iv}~$\lambda\lambda 1393, 1402$}
\newcommand{\NaIdblt}{{\rm Na}\kern 0.1em{\sc i}~$\lambda\lambda 5890, 5896$}
\newcommand{\PV}{\hbox{{\rm P}\kern 0.1em{\sc v}}}
\newcommand{\AlI}{\hbox{{\rm Al}\kern 0.1em{\sc i}}}
\newcommand{\AlII}{\hbox{{\rm Al}\kern 0.1em{\sc ii}}}
\newcommand{\AlIII}{{\hbox{\rm Al}\kern 0.1em{\sc iii}}}
\newcommand{\CaII}{\hbox{{\rm Ca}\kern 0.1em{\sc ii}}}
\newcommand{\CII}{\hbox{{\rm C}\kern 0.1em{\sc ii}}}
\newcommand{\CIIe}{\hbox{{\rm C$^{\ast}$}\kern 0.1em{\sc ii}}}
\newcommand{\CIII}{\hbox{{\rm C}\kern 0.1em{\sc iii}}}
\newcommand{\CIV}{\hbox{{\rm C}\kern 0.1em{\sc iv}}}
\newcommand{\CV}{\hbox{{\rm C}\kern 0.1em{\sc v}}}
\newcommand{\HI}{\hbox{{\rm H}\kern 0.1em{\sc i}}}
\newcommand{\HII}{\hbox{{\rm H}\kern 0.1em{\sc ii}}}
\newcommand{\Ha}{\hbox{{\rm H}\kern 0.1em$\alpha$}}
\newcommand{\Lya}{\hbox{{\rm Ly}\kern 0.1em$\alpha$}}
\newcommand{\Lyb}{\hbox{{\rm Ly}\kern 0.1em$\beta$}}
\newcommand{\Lyg}{\hbox{{\rm Ly}\kern 0.1em$\gamma$}}
\newcommand{\Lyd}{\hbox{{\rm Ly}\kern 0.1em$\delta$}}
\newcommand{\Lye}{\hbox{{\rm Ly}\kern 0.1em$\epsilon$}}
\newcommand{\Lyphi}{\hbox{{\rm Ly}\kern 0.1em$\phi$}}
\newcommand{\Lyfive}{\hbox{{\rm Ly}\kern 0.1em$5$}}
\newcommand{\Lysix}{\hbox{{\rm Ly}\kern 0.1em$6$}}
\newcommand{\Lyseven}{\hbox{{\rm Ly}\kern 0.1em$7$}}
\newcommand{\Lyeight}{\hbox{{\rm Ly}\kern 0.1em$8$}}
\newcommand{\Lynine}{\hbox{{\rm Ly}\kern 0.1em$9$}}
\newcommand{\Lyten}{\hbox{{\rm Ly}\kern 0.1em$10$}}
\newcommand{\Lyeleven}{\hbox{{\rm Ly}\kern 0.1em$11$}}
\newcommand{\HeI}{\hbox{{\rm He}\kern 0.1em{\sc i}}}
\newcommand{\HeII}{\hbox{{\rm He}\kern 0.1em{\sc ii}}}
\newcommand{\FeI}{\hbox{{\rm Fe}\kern 0.1em{\sc i}}}
\newcommand{\FeII}{\hbox{{\rm Fe}\kern 0.1em{\sc ii}}}
\newcommand{\FeIII}{\hbox{{\rm Fe}\kern 0.1em{\sc iii}}}
\newcommand{\MnII}{\hbox{{\rm Mn}\kern 0.1em{\sc ii}}}
\newcommand{\MgI}{\hbox{{\rm Mg}\kern 0.1em{\sc i}}}
\newcommand{\MgII}{\hbox{{\rm Mg}\kern 0.1em{\sc ii}}}
\newcommand{\MgIII}{\hbox{{\rm Mg}\kern 0.1em{\sc iii}}}
\newcommand{\NI}{\hbox{{\rm N}\kern 0.1em{\sc i}}}
\newcommand{\NII}{\hbox{{\rm N}\kern 0.1em{\sc ii}}}
\newcommand{\NIII}{\hbox{{\rm N}\kern 0.1em{\sc iii}}}
\newcommand{\NV}{\hbox{{\rm N}\kern 0.1em{\sc v}}}
\newcommand{\OVI}{\hbox{{\rm O}\kern 0.1em{\sc vi}}}
\newcommand{\OI}{\hbox{{\rm O}\kern 0.1em{\sc i}}}
\newcommand{\OII}{\hbox{[{\rm O}\kern 0.1em{\sc ii}]}}
\newcommand{\OIV}{\hbox{{\rm O}\kern 0.1em{\sc iv}]}}
\newcommand{\SI}{{\rm S}\kern 0.1em{\sc i}}
\newcommand{\SIV}{{\rm S}\kern 0.1em{\sc iv}}
\newcommand{\SVI}{{\rm S}\kern 0.1em{\sc vi}}
\newcommand{\SiI}{\hbox{{\rm Si}\kern 0.1em{\sc i}}}
\newcommand{\SiII}{\hbox{{\rm Si}\kern 0.1em{\sc ii}}}
\newcommand{\SiIII}{\hbox{{\rm Si}\kern 0.1em{\sc iii}}}
\newcommand{\SiIV}{\hbox{{\rm Si}\kern 0.1em{\sc iv}}}
\newcommand{\SII}{\hbox{{\rm S}\kern 0.1em{\sc ii}}}
\newcommand{\SIII}{\hbox{{\rm S}\kern 0.1em{\sc iii}}}
\newcommand{\NaI}{\hbox{{\rm Na}\kern 0.1em{\sc i}}}
\newcommand{\TiII}{\hbox{{\rm Ti}\kern 0.1em{\sc ii}}}
\newcommand{\kms}{\hbox{km~s$^{-1}$}}
\newcommand{\cmsq}{\hbox{cm$^{-2}$}}

\newcommand{\kmsM}{\hbox{km~s$^{-1}$~Mpc$^{-1}$}}
\newcommand{\Mc}{\hbox{Mpc$^{-3}$}}
\newcommand{\ergsec}{\hbox{ergs~s$^{-1}$}}
\newcommand{\msun}{M$_{\odot}$}

\tighten
\begin{document}
 
 
\lefthead{BOND ET~AL.}
\righthead{SOURCE OF STRONGEST {\MgII} ABSORBERS}


\title{High--Redshift Superwinds as the Source of the Strongest {\MgII} Absorbers:  A Feasibility Analysis\altaffilmark{1}}

\author{Nicholas~A.~Bond, Christopher~W.~Churchill\altaffilmark{2}, Jane~C.~Charlton\altaffilmark{3}} 
\affil{Department of Astronomy and Astrophysics \\
       The Pennsylvania State University \\ University Park, PA 16802
       \\ {\it bond, cwc, charlton@astro.psu.edu}}
\and 
\author{Steven S. Vogt\altaffilmark{2}}
\affil{UCO/Lick Observatories \\
       Board of Studies in Astronomy and Astrophysics \\
       University of California, Santa Cruz, CA 96054 \\
       {\it vogt@ucolick.org}}

\altaffiltext{1}{Based in part on observations obtained at the
W.~M. Keck Observatory, which is operated as a scientific partnership
among Caltech, the University of California, and NASA. The Observatory
was made possible by the generous financial support of the W.~M. Keck
Foundation.}
\altaffiltext{2}{Visiting Astronomer, W.~M. Keck Observatory}
\altaffiltext{3}{Center for Gravitational Physics and Geometry}

\begin{abstract}

We present HIRES/Keck profiles of four extremely strong ($W_r >
1.8$~{\AA}) {\MgII} absorbers at $1<z<2$.  The profiles display a
common kinematic structure, having a sharp drop in optical depth near
the center of the profile and strong, black--bottomed absorption on
either side.  This ``symmetric-inverted'' structure, with a velocity
spread of several hundred kilometers per second, is suggestive of
superwinds arising in actively star--forming galaxies.
Low--ionization absorption of similar strength has been observed in
local star--forming galaxies.
        
The {\MgII} absorbers with $W_r > 1.8$~{\AA} evolve away from $z=2$ to
the present.  We propose that a substantial fraction of these very
strong absorbers are due to superwinds and that their evolution is
related to the redshift evolution of star--forming galaxies.  Based on
the observed redshift number density of $W_r > 1.8$~{\AA} {\MgII}
absorbers at $1 < z < 2$, we explore whether it is realistic that
superwinds from starbursting galaxies could give rise to these
absorbers.  Finally, we do an analysis of the superwind connection to
damped {\Lya} absorbers (DLAs).  DLAs and superwinds evolve
differently and usually have different kinematic structure, indicating
that superwinds probably do not give rise to the majority of DLAs.

\end{abstract}



\section{Introduction}

The very strong ($W_r(2796) > 1.8$~{\AA}), $z>1$ {\MgII} absorbers are
of great interest in determining the nature of the
high--column--density and/or high--velocity material present in the
early universe.  At low redshift, very strong {\MgII} systems evolve
away as ($1+z$)$^{2.24}$ (\cite{Steidel92}).  This evolution parallels
the star formation history of the universe (e.g. \cite{Yan99}),
suggesting a link between {\MgII} absorption and star formation
activity (e.g.\ \cite{Guillemin97}, \cite{Churchill99}).  The
disappearance of these absorbers at low redshifts also suggests that
the gas giving rise to the absorbers has dispersed and/or changed
state with time.  Both of these factors suggest a connection between
very strong {\MgII} absorbers and superwinds in starburst galaxies.

Star formation in the nucleus of a starburst galaxy is thought to give
rise to a superbubble, a hollow shell of swept--up ISM gas (e.g.\
\cite{MM88}).  If this superbubble expands sufficiently to escape the
galaxy's potential well, it can ``blow out'' into the halo of the
galaxy.  At this point, Rayleigh--Taylor instabilities would cause the
shell to fragment and the superbubble would begin to vent hot material
into the halo of the starbursting galaxy, forming a bipolar,
weakly--collimated outflow (e.g., {\cite{Tomisaka93}};
{\cite{Suchkov94}}; {\cite{Strickland98}}; {\cite{TT98}}).

The central region of the superwind/superbubble is very hot ($T \sim
10^7$~K) and gives rise to X--ray emission.  There may be a conduction
zone between the shell and the central region with coronal gas ($T
\sim 10^5$~K) that exhibits absorption in high--ionization UV lines
(e.g. {\CIV} and {\OVI}).  The superwind/superbubble shell, however,
is cold ($T \sim 10^4$~K) and exhibits low--ionization absorption in
NaI~$\lambda\lambda 5890,5896$ (\cite{HeckmanNaD}), indicating that
{\MgII} absorption would also be present.  Some material associated
with the cold phase emits in {\Ha} because of the ionizing flux coming
from the young stars.  Rayleigh--Taylor instabilities would result in
clumping of the cold material and a line of sight through these clumps
would be expected to show kinematically complex absorption.

Heckman \etal (2001\nocite{Heckman1705}) observe the absorption
spectrum of a dwarf galaxy, NGC $1705$, at high resolution with {\it
FUSE}.  There is evidence for a $100$~{\kms} outflow with both cold
and warm phases of gas.  Kinematically, the warm phase seems
consistent with an expanding superbubble (the beginnings of a
superwind).  Rather than simply having the warm phase in a conduction
zone, however, Heckman \etal (2001\nocite{Heckman1705}) hypothesize
that hot gas is blowing out through the shell and mixing with the cold
gas.  This process creates warm gas of $\sim 10^5$~K gas that gives
rise to {\OVI} absorption.  The cold phase, on the other hand, is
expected to arise in the shell.  The absorption lines near the
systemic velocity of NGC $1705$ show $\sim 100$~{\kms} spreads.

In this paper, we argue that a substantial fraction of the
$W_r>1.8$~{\AA} {\MgII} absorption systems are produced by superwinds.
In \S~\ref{sec:data}, we present several HIRES/Keck absorption
profiles of very strong, $z>1$ {\MgII} absorbers.  In
\S~\ref{sec:hypothesis}, we establish the basis for our hypothesis
that superwinds give rise to {\MgII} absorption systems like the ones
in our sample.  In \S~\ref{sec:results}, we calculate the expected
redshift number density, $N(z)$ of these absorbers if they are caused
by superwinds.  In \S~\ref{sec:dndzlocal}, we assess the uncertainty of
the parameters going into the $N(z)$ calculation.  In
\S~\ref{sec:discussion}, we briefly discuss the implications for
Lyman break galaxies and damped {\Lya} absorbers (DLAs).  Finally, in
\S~\ref{sec:conclude}, we summarize our results and conclusions.

\section{Data and Analysis}
\label{sec:data}

\subsection{Sample Selection and Observations}

During the observing runs for a larger program to study the kinematics
of {\MgII} absorbers at $0.5 < z < 1.5$ (\cite{csv96};
\cite{kinmods}; \cite{weak}; \cite{cv01}), several very large equivalent width 
systems ($W_r > 1.8$~{\AA}) were observed at $z>1$.  The goal was to
obtain a small sample of kinematic data from systems belonging to the
strongly evolving population of $W_r>1.8$~{\AA} {\MgII} absorbers
documented by Steidel \& Sargent (1992\nocite{Steidel92}).

Four {\MgII} systems were targeted in three quasars.  The systems,
shown in Figure~\ref{fig:strong} are at $z=1.1745$ toward
Q~$0450-132$, $z=1.3201$ and $z=1.5541$ toward Q~$1213-003$, and
$z=1.7948$ toward B2~$1225+317$.  The latter system was studied in
detail at high spectral resolution by Bechtold \etal
(1987\nocite{Bechtold87}).

The four quasars were observed with the HIRES spectrometer
(\cite{Vogt94}) on the Keck~I telescope on the night of 24 January
1995 UT.  HIRES was configured in first--order using the
$0.861$~{\arcsec} slit width and either the $7$~{\arcsec} or $14$~{\arcsec}
decker.  Because of the first--order HIRES format, there are small
gaps in the spectral coverage red-ward of $5100$~{\AA}, where the free
spectral range exceeds the width of the $2048 \times 2048$ Tektronix
CCD.  The spectral resolution is $R = 45,000$ ($\simeq 6.6$~{\kms}).

The journal of observations is presented in
Table~\ref{tab:obsjournal}, which lists, for each quasar, the $V$
magnitude, emission redshift, observation date, total integration
time, wavelength coverage (not including gaps above $5100$~{\AA}),
decker size for sky subtraction, and high pass filter for 2nd--order
blocking.  Three separate, consecutive integrations were obtained for
each quasar spectrum, bracketed by Th--Ar calibration lamps.

\subsection{Data Reduction and Analysis}

The HIRES data were reduced with the IRAF\footnote{IRAF is distributed
by the National Optical Astronomy Observatories, which are operated by
AURA, Inc., under contract to the NSF.}  {\sc Apextract\/} package
(V2.10.3) for echelle data.  Each observed data frame was overscan
subtracted, bias frame corrected, scattered light corrected, and
flat-fielded in the standard fashion.  The data have been calibrated to
vacuum wavelengths and converted to heliocentric velocity.  The
unfluxed continuum was normalized using the methods described by
Sembach \& Savage (1992\nocite{semsav92}).  Further details of the
data reduction can be found in Churchill (1997\nocite{thesis});
Churchill \etal (1999\nocite{weak}), Churchill \etal
(2000\nocite{archive1}), and Churchill \& Vogt (2001\nocite{cv01}).

Equivalent widths and equivalent width limits were computed using the
methods of Lanzetta, Wolfe, \& Turnshek (1987\nocite{lwt87}).  We
focus on the {\MgII}, {\FeII}, {\MgI}, and {\MnII} transitions because
they are represented for all four systems (due to a combination of
their redshifted wavelengths and the wavelength coverage of the HIRES
echelle).  The apparent optical depth (AOD, \cite{savsems91}) column
densities, $N_{a}$, were also computed for each atom/ion.  The $N_{a}$
values were determined by combining information from all unsaturated
transitions as described in the appendix of Churchill \& Vogt
(2001\nocite{cv01}).  When limits are placed, they are simply the
measured $N_{a}$ of the weakest transition (i.e. no correction for
unresolved saturation is applied; {\cite{Jenkins96}}).

In Table~\ref{tab:properties}, we list the equivalent widths or their
limits ($3~\sigma$), and the adopted AOD column densities for each
system.  We also present the total velocity width, using the velocity
range over which the equivalent widths were computed.

\section{Superwind Hypothesis}
\label{sec:hypothesis}

\begin{figure*}
\figurenum{1}
\plotfiddle{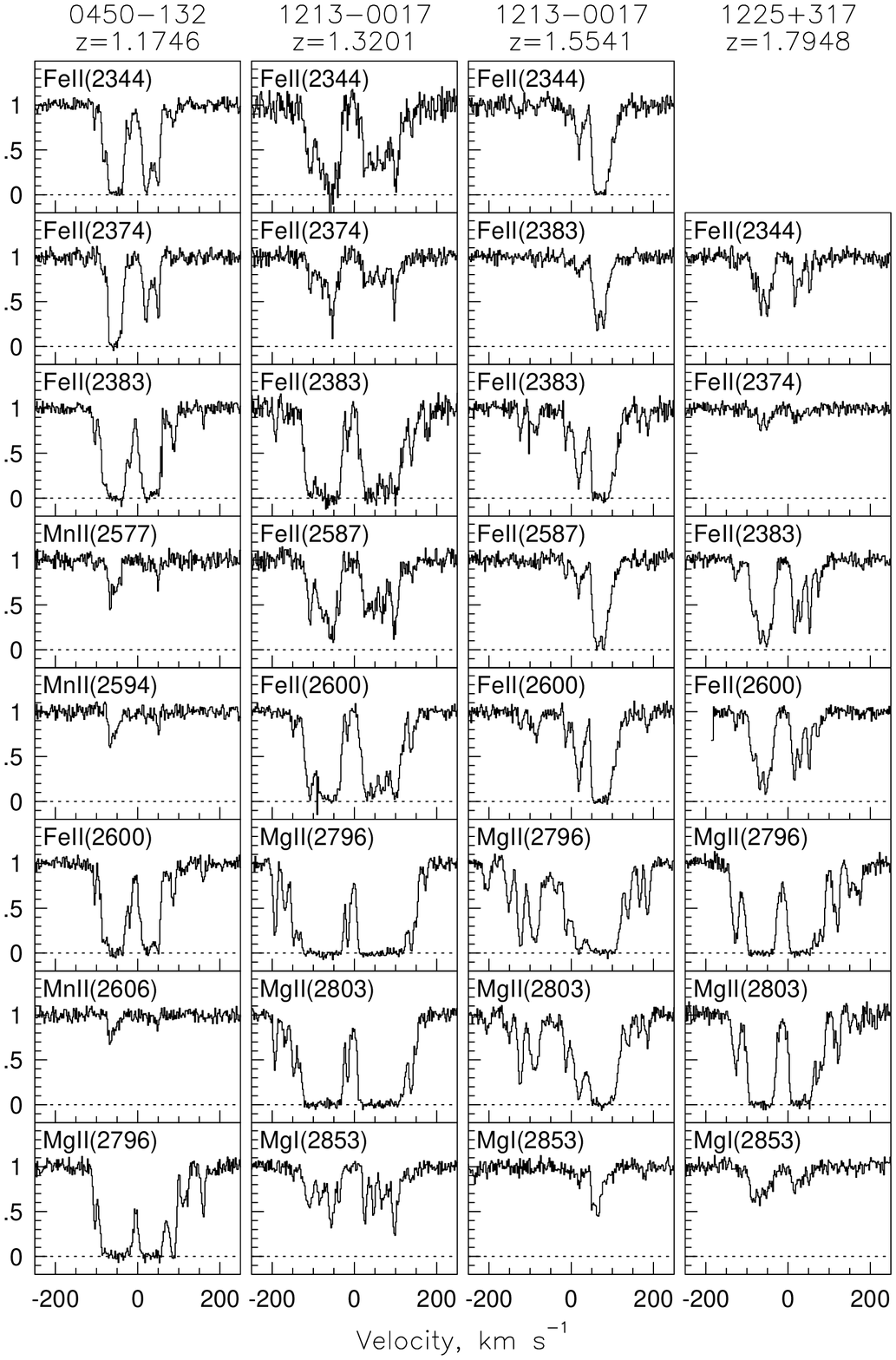}{6.in}{0}{80}{80}{-270}{-50}
\caption{
Four $W_r>1.8$~{\AA} {\MgII} absorption systems at $1 < z < 2$.  The
{\MgIIdblt} doublet and the corresponding detected {\FeII}, {\MgI},
and {\MnII} transitions are plotted, aligned in rest--frame velocity.
The {\MgII}~2803 transition was not covered in the spectrum of
Q~$0450-132$.  Data were obtained at $R \sim 6.6$~{\kms} with the
HIRES spectrograph on the Keck I telescope.  }
\label{fig:strong}
\end{figure*}

We hypothesize that the profiles presented in Figure~\ref{fig:strong}
arise in a quasar spectrum when the line of sight passes through the
outflowing gas of a superwind.  There are several distinctive features
of the profiles that can be easily explained with this hypothesis: 1)
a narrow ``inversion'' (an anomalous region of low optical depth) at
kinematic center, 2) many absorption components spread out over a
large velocity range ($\sim 250$~{\kms}), and 3) a variation in
cloud--to--cloud abundance patterns.

The most prominent feature is the inversion at the center of the
profile.  This type of feature implies that the absorbing material is
either expanding outward or falling inward, with little or no material
moving at the systemic velocity of the absorbing medium.  The outflows
of superwinds behave in this way, typically forming a double--cone
structure that is symmetric about the minor axis of the host galaxy
(e.g. \cite{Heckman90}).  If a quasar line of sight were to pass
through both cones, one would see blue--shifted absorption from one
cone and red--shifted absorption from the other cone.  If the galaxy
inclination was very close to face--on, there might be absorption from
the disk at the systemic velocity, but face--on orientations are rare
and the disk absorption would be blue-- or red--shifted at other
inclinations.  The inversion could also be created by passing through
a single cone of a superwind.  Most of the {\MgII} absorption is
created in the cold shell along the edges of the outflow.  The center
of the wind, which would correspond to the center of the profile, is
mostly filled with hot ($\sim 10^6$~K) gas and would not create much
{\MgII} absorption.

A typical velocity spread for these systems is $\sim 250$~{\kms}.
This is consistent with the typical superwind velocities, which are
observed to be between $100$ and $1000$~{\kms}
(e.g. \cite{HeckmanNaD}).  In fact, Heckman \etal
(2001\nocite{Heckman01}, in prep) have observed the {\MgII} absorption
from a known superwind, NGC$1705$.  They use the starburst region
itself as a continuum and are looking through only one side of the
wind, so we expect that they should measure about half the velocity
spread of the very strong {\MgII} absorbers in our sample.  They find
$\Delta v \sim 150$~{\kms}, a spread consistent with our data.

Another feature of the profiles that is immediately obvious is the
many absorption components.  Weak transitions, like those of {\MgI}
and {\FeII}, clearly show that the saturated troughs seen in {\MgII}
are made up of many components in velocity space.  In the context of
the superwind model, these clouds most likely correspond to the
fragmented pieces of the entrained material.  These fragments are
produced by Rayleigh--Taylor instabilities in the shell of the initial
superbubble when it blows out of the host galaxy.  The shell pieces
are then entrained and become dispersed along the edges of the
weakly--collimated outflow, producing the multi--component absorption
troughs.

The final feature that stands out in Figure~\ref{fig:strong},
particularly in the $z=1.1746$ system in Q$0450-132$, is the variation
in cloud--to--cloud line ratios.  For example, in the Q$0450-132$
system, the ratio of the line strengths of {\FeII} and {\MnII} varies
dramatically across the profile.  Since these species have similar
recombination rates and photoionization cross sections, this indicates
that the abundance patterns of the gas varies from cloud to cloud.
This kind of behavior might be expected in a superwind if the
fragments of entrained material causing the absorption originated in
different regions of the ISM of the pre--starburst galaxy, each
populated with different abundance patterns.

\subsection{Alternative Explanations}
\label{sec:alternatives}

An alternative hypothesis as to the origin of the very strong {\MgII}
absorbers was put forth by Danks \etal (2001\nocite{Danks01}).  Based
on absorption profiles from the regions surrounding the Carinae
Nebula, they suggest that some very strong {\MgII} absorption could be
due to interstellar gas in the vicinity of active star formation
regions.  Unlike in the superwinds, however, the gas in these nebulae
is very cold--there is very little warm gas.  Therefore, observations
of high--ionization transitions, like {\CIV}, would help confirm or
refute this hypothesis.  In addition, these nebulae would be expected to
exhibit variability, as seen in the Carinae Nebula, over timescales of
a few years.  However, in the superwind scenario, even if the clouds
were as small as $1$~pc in size and moving at velocities of a few
hundred {\kms}, we would expect observable variability over timescales
of no less than $10^3$ or $10^4$ years.  Multi--epoch observations of
the very strong {\MgII} systems are required to determine whether
interstellar gas could be the source of the absorption.

A further possibility for the origin of these absorbers is a pair of
galaxies in a group or cluster or a large galaxy and its satellite
(\cite{Churchillconf}).  Although there may be enough of these pairs
to explain the frequency of the ``symmetric--inverted'' systems, they
do not provide as natural an explanation for the kinematics.  If this
absorption was due solely to galaxies, one would expect to observe
many more systems with single saturated troughs surrounded by a
significant number of small outlying clouds.  Also, the splitting of the
troughs in velocity space shows remarkably little variation.  One
would expect pairs of galaxies to show a larger range of velocity
differences.

\subsection{Q$1213-0017$ Field}

The field of the quasar Q$1213-0017$ has been the target of extensive
ground--based and {\it HST} imaging in the optical and near--IR
(\cite{Liu00}).  To date, four galaxies have been found within $\sim
1$~Mpc of the quasar at $z \sim 1.3$, the redshift of one of the
intervening strong {\MgII} absorbers in the quasar spectrum.  The
closest of these galaxies is at an impact parameter{\footnote{In this
paper, we use $H_0 = 75$~{\kmsM}, $q_0 = 0.5$ cosmology.}} of
$195$~kpc and is unlikely to be responsible for the {\MgII} absorption
at that redshift.  However, the abundance of bright galaxies at this
redshift is indicative of a cluster and there are a few fainter
objects in the field that could not be studied spectroscopically.  If
these objects are at $z \sim 1.3$, they could be causing the
absorption.

In the superwind scenario, we would expect the majority of
$W_r>1.8$~{\AA} {\MgII} absorption systems to arise in relatively
low--luminosity galaxies below the detection threshold of the
Q$1213-0017$ images.  For example, NGC$1705$ would not be detected in
the images of Liu \etal (2000\nocite{Liu00}).  Because there is a
cluster near the redshift of the absorption system it is not
unreasonable to think that the line of sight is passing through the
wind of a low--luminosity starburst or a relic outflow.

\section{Feasibility Analysis}
\label{sec:results}

We wish to determine the feasibility of our hypothesis that superwinds
give rise to these very strong ``symmetric--inverted'' {\MgII} systems
at high redshift.  To do this, we first estimate the observed redshift
number density, $N(z)$, of {\MgII} absorbers with $W_r > 1.8$~{\AA} at
$\left<z\right> \sim 1.3$ from quasar absorption line surveys.  Then,
we explore the plausible range of $N(z)$ of these absorbers assuming
they are superwinds associated with starburst galaxies.  In performing
this calculation, we parameterize the physical properties of
superwinds and find the parameter space consistent with the observed
$N(z)$ of very strong symmetric--inverted {\MgII} systems.

\subsection{Observed $N(z)$ of Very Strong {\MgII} Systems}
\label{sec:dndzobs}

As can be seen in Figure~$2$ of Steidel \& Sargent
(1992\nocite{Steidel92}), the observed $N(z)$ of $W_r>1$~{\AA} {\MgII}
absorption systems is $\sim 0.3$.  The statistics were not sufficient
to determine an $N(z)$ for a higher minimum $W_r$, so we can only
place a limit on the $N(z)$ of $W_r>1.8$~{\AA} systems, $N(z)<0.3$.
We can approximate the actual value by making an extrapolation of the
curve.  After doing this, we get $N(z) \sim 0.15$ and use this value
in the calculations below.  Although the exact value of $N(z)$ is
uncertain, the following calculation is rough and nature and a
slightly different value will not affect our conclusions.

\subsection{Estimated $N(z)$ of Absorbing Superwinds}
\label{sec:dndzcalc}

For an absorber with number density $n(z)$ and cross section $\sigma (z)$,
the number of absorbers per unit redshift path is
\begin{equation}
N(z) = \frac{c~n(z)~\sigma (z)}{H_0}~(1 + z)~(1 + 2~q_0~z)^{1/2}.
\label{eq:dndz}
\end{equation}
The cross section and density cannot be determined independently.
Assuming the absorption is associated with a population of galaxies at
$z=1.3$, the product of the two can be obtained by integrating
\begin{equation}
n~\sigma = C_f~C_t~\int _{L_{min}}^{\infty} \phi(L/L^*)~\pi~{R(L/L^*)}^2~dL,
\label{eq:nsigint}
\end{equation}
where $C_f$ is the covering factor, $C_t$ is a time-scale factor
explained below, and $L_{min}$ is the minimum {\Ha} luminosity that a
galaxy can have and still give rise to symmetric--inverted {\MgII}
absorption.  We used $\phi(L/L^*) =
\Phi^*~(L/L^*)^{\alpha}~e^{-L/L^*}$, a Schecter formulation of a
galaxy luminosity function (\cite{Schecter76}), and $R(L/L^*) =
R^*~(L/L^*)^{\beta}$, a Holmberg formulation of a radius--luminosity
relationship (\cite{Holmberg75}), where $R^*$ is the radius out to
which one gets very strong absorption in an $L^*$ galaxy.

When the integration is carried out, one obtains
\begin{equation}
n~\sigma = C_f~C_t~\Phi^*~\pi~R_*^2~\Gamma(\alpha + 2~\beta + 1,L_{min}/L^*),
\label{eq:nsig}
\end{equation}
where $\Gamma$ is the incomplete gamma function.  The integral is
sensitive to the choice of $L_{min}$, especially for small $\beta$, so
we explore a range of $L_{min}$.  The incomplete gamma function
was calculated using Mathematica (v. 4.0.1.0).

Strong {\Ha} emission is an indicator of high star--formation rates
and starburst activity.  Therefore, the luminosity function of
starburst galaxies was assumed to follow an {\Ha} luminosity function.
At $z=1.3$, Yan \etal (1999\nocite{Yan99}) find $\alpha = -1.35$,
$\Phi^* = 1.7 \times 10^{-3}$~{\Mc}, and $L^* = 7 \times
10^{42}$~{\ergsec}.

In Equation~\ref{eq:nsigint} and \ref{eq:nsig}, the factor $C_t$
arises because the cold gas causing the {\MgII} absorption is expected
to remain after the outflow has ceased and the galaxy is no longer
visible in {\Ha}.  We parameterize this as $C_t = \tau /t_{{\Ha}}$,
where $\tau$ is the time over which one can observe very strong
absorption from the superwind and $t_{\Ha}$ is the ratio of the
lifetime of the symmetric--inverted systems to the lifetime of the
{\Ha} luminosity.

Plugging Equation~\ref{eq:nsig} into Equation~\ref{eq:dndz} and using
$N(z)=0.15$ and $C_f=1$, we obtain a family of $C_t$ vs.\
$L_{min}/L^*$ curves, which we present in Figure~{\ref{fig:dndz}}.  We
place $C_t$ and $L_{min}$ on the axes because they are the most
uncertain parameters (see \S~\ref{sec:dndzlocal}).  If values are
chosen for $R^*$, $\alpha$, and $\beta$, the $C_t$ and $L_{min}$
values necessary to produce $N(z)=0.15$ can be read from the curves.
Separate panels are shown for $R^*=2$, $5$, $10$, and $20$~kpc and the
curves shown are for $\beta = 0$, $0.05$, $0.1$, and $0.15$.  Note
that a change in $\beta$ by an amount $\Delta \beta$ is equivalent to
a change in $\alpha$ by an amount $2\Delta \beta$.

\section{Constraining Parameter Space}
\label{sec:dndzlocal}

Theoretical models and observations of local superwinds can provide us
with reasonable approximations of $\beta$, $R^*$, $C_t$, and
$L_{min}$ for comparison to our results in Figure~\ref{fig:dndz}.

\subsection{Timescale}
\label{sec:timescale}

The {\Ha} luminosity will persist as long as there is a steady flow of
ionizing photons from the nucleus of the starburst.  Therefore, the
{\Ha} luminosity should last as long as the O and B stars producing
the starburst and driving the superwind ($\sim 10^7$~years).  The
timescale for {\MgII} absorption must be at least this long because it
is observed in the post--starburst dwarf, NGC$1705$.  The
low--ionization gas will certainly persist for longer than this, but
diffusion will eventually cause the material to spread out enough that
the very strong {\MgII} systems are no longer seen.  The timescale for
this is not known exactly, but it is probably at least a few times the
age of the starburst.  The reason that we have used $C_t$ as the
ordinate of Figure~\ref{fig:dndz} is because it is one of the most
uncertain parameters. 

\subsection{Minimum Luminosity}
\label{sec:lmin}

We can place a constraint on $L_{min}$ by again considering NGC$1705$,
which has an {\Ha} luminosity of $\log (L_{\Ha}/L^*)=-2.7$
(\cite{Marlow97}).  Though NGC$1705$ is a starburst known to produce
{\MgII} absorption over a large velocity spread, it has not been
observed along a background quasar line of sight and we cannot be
certain that a black--bottomed profile with a central inversion would
be produced.  Therefore, we explore a range of $-3.5 < \log L_{min} <
-1.5$, allowing for uncertainty in either direction.  In
Figure~\ref{fig:dndz}, the curves rise dramatically (and require large
$C_t$) at high $L_{min}$, so we expect that, to produce $N(z)=0.15$,
$\log L_{min}$ must be less than about $-2.5$.

\subsection{Radius--Luminosity Slope}
\label{sec:beta}

The similarity solution of Mac Low \& McCray (1988\nocite{MM88})
predicts $R \propto L^{0.2}$ for an expanding superbubble, where $L$
is the total luminosity of the early--type stars driving the bubble.
Most of this luminosity will be hydrogen--ionizing, so one would
expect $R \propto L_{H{\alpha}}^{ 0.2}$ as well.  The later stages of
the superwind (post--blowout), however, cannot be treated as an
expanding superbubble.  At this point in the evolution of the
superwind, the outflowing material is no longer being accelerated away
from the galaxy but is, rather, being decelerated by ram pressure from
the intergalactic medium (IGM) and by the galaxy potential well.
Post--blowout outflows in more massive (and therefore more luminous)
galaxies will have higher--velocity outflows, resulting in greater ram
pressure from the IGM.  More massive galaxies would also experience
greater gravitational deceleration.  Both of these effects act to
decrease the radius--luminosity slope to $\beta < 0.2$.  We only plot
curves for $0 \le \beta \le 0.15$ in Figure~\ref{fig:dndz} because the
material will probably disperse before it has been significantly
decelerated.  Lower values of $\beta$ produce more absorbers because
they imply that the Holmberg relation is flat ($\beta \sim 0$) and the
total cross section is being dominated by the large numbers of dwarf
galaxies at the faint end of the luminosity function.

\subsection{Radius--Luminosity Normalization}
\label{sec:rstar}

The {\NaI} absorption radii of far--IR--bright starburst galaxies can
be used to estimate plausible values of $R^*$ for {\MgII}.  Absorption
in {\NaI} is known to occur out to radii of $5$~kpc with $C_f=1$
(\cite{HeckmanNaD}).  The radii are lower limits because the
background starlight becomes too faint beyond this point and because
{\MgII} may exist out to a larger radius than {\NaI}.  If we are
conservative and take $R^*=5$~kpc, we require a low $L_{min}$ and/or a
high $C_t$.  There is reason to believe, however, that the absorption
will be seen at larger radii.  In a low--resolution spectrum, Norman
\etal (1996\nocite{Norman96}) detected $W_r=1.7$~{\AA} {\MgII}
absorption at an impact parameter of $26$~kpc from a starburst galaxy,
but it is not known whether the system has the type of kinematic
structure we are observing at high redshift.  Also, {\Ha} emission has
been seen out to an impact parameter of $11$~kpc in M$82$
(\cite{Devine99}), indicating the presence of cold material out to
that radius.  Radii of $10$~kpc or greater will produce the observed
$N(z)$ for moderate $L_{min}$ and $C_t$ factors of a few.

\subsection{Summary}
\label{sec:summary}

We expect $1 \la C_t \la 5$ based on the speed and duration of the
superwind.  Although we can't place a useful constraint on $L_{min}$
from physical considerations, we argue $0 \la \beta < 0.2$ by
considering the pressure mechanisms acting on the expanding
superbubble.  Observations of local starburst galaxies tell us that $5
\la R^* \la 25$.

\begin{figure*}
\figurenum{2}
\plotfiddle{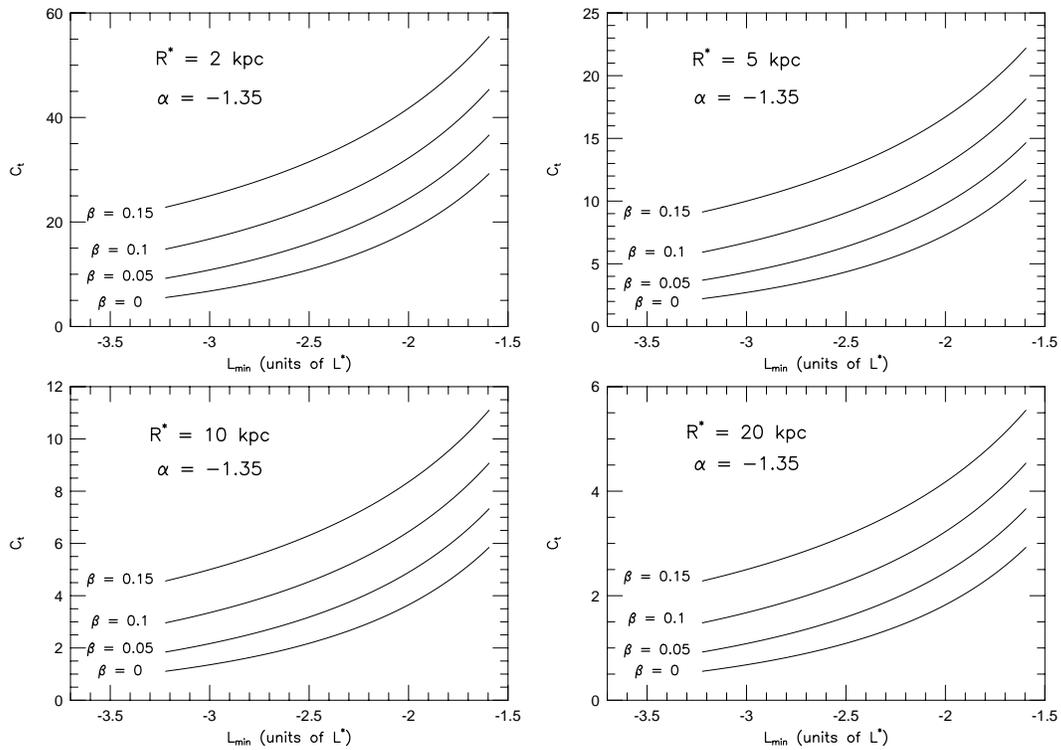}{6.in}{0}{60}{60}{-250}{0}
\caption{
Plots of the parameters required to produce the observed number of {\MgII}
absorbers per unit redshift ($N(z)$) with $W_r>1.8$~{\AA}.  Each curve
represents the parameters ($C_t$ and $L_{min}$) needed to produce the
observed $N(z)=0.16$ for a given $R^*$ and $\beta$.  The vertical axis
gives the absorption timescale factor ($C_t$) and the horizontal axis,
the minimum {\Ha} luminosity ($L_{min}$) for which very strong {\MgII}
absorption would be observed.  Each plot assumes a different scale
radius ($R^*$) for the radius--luminosity relation,
$R=R^*~(L/L^*)^{\beta}$, and curves are presented for different values
of $\beta$ as labelled.  }
\label{fig:dndz}
\end{figure*}

Using these physical arguments, we can further constrain parameter
space by examining Figure~\ref{fig:dndz}.  In order to obtain
$N(z)=0.15$ for reasonable values of $C_t$, we expect that $R^* \sim
10$~kpc and $\log L_{min} \la -2.5$.  The calculation is not
particularly sensitive to $\beta$ within the chosen range, so we can't
place any further constraint on it from Figure~\ref{fig:dndz}.  The
timescale factor ($C_t$) cannot be constrained from
Figure~\ref{fig:dndz} either.  

It should be recognized that the calculations presented here are very
rough.  We can only determine that the parameters needed to produce
the observed $N(z)$ of $W_r>1.8$~{\AA} {\MgII} absorption systems are
consistent with the expected properties of superwinds at $z \sim 1.3$.

\section{Discussion}
\label{sec:discussion}

\subsection{$N(z)$ of Lyman break Galaxies}
\label{sec:lymbreak}

High--redshift analogs of local superwinds, Lyman break galaxies,
have been seen by Steidel et al. (1996\nocite{Steidel96}) at $z
\sim 3$.  A rest--frame ultraviolet spectrum of a lensed Lyman break galaxy,
MS 1512--cB58, shows evidence for an outflow of $\sim 200$~{\kms} and a
star formation rate of $\sim 40$~{\msun}~yr$^{-1}$.  These properties
are similar to those we expect for the objects giving rise to the very
strong {\MgII} absorbers, indicating that {\MgII} absorption could be
used as a tracer of Lyman break galaxies at $z \sim 3$. 

Using equation~\ref{eq:dndz} and existing data on Lyman break
galaxies, we can estimate the expected redshift number density of
{\MgII} absorbers from Lyman break galaxies.  If we assume that these
objects are like the $z \sim 1.3$ very strong {\MgII} absorbers and
use $R^*=5$~kpc (see Section~\ref{sec:dndzlocal}) as the typical
absorption radius, we get $\sigma \simeq 80$~kpc$^2$.  The number
density of Lyman break galaxies is $n \ga 0.02$~Mpc$^{-3}$
(\cite{Adelberger98}).  This is a lower limit because it only takes
into account those Lyman break galaxies that have been detected using
photometric methods.  Using equation~\ref{eq:dndz}, we get $N(z)_{LBG}
\ga 0.05$ at $z=3$.

\subsection{Superwind Contribution to Damped Lyman--Alpha Absorbers}
\label{sec:DLA}

The idea that superwinds seen in absorption might constitute a
non--negligible fraction of the damped {\Lya} absorbers (DLAs, $\log
N(\HI)>20.3$~{\cmsq}) in quasar spectra was first proposed by Nulsen
\etal (1998\nocite{Nulsen98}).  Based on a rough model of a weakly
collimated, biconical outflow, they determined that superwinds could
give rise to the majority of DLAs at all redshifts.  We assess whether
the majority of symmetric--inverted ($W_r>1.8$~{\AA}) {\MgII} profiles
that we attribute to superwinds could also be DLAs.

Of the absorbers shown in Figure~{\ref{fig:strong}}, two have measured
{\HI} column densities.  The system in Q$1213-0017$ at $z=1.5541$ was
found not to be a DLA (\cite{Rao00}) from analysis of a {\it HST}/FOS
spectrum.  For the system in Q$1225+317$ at $z=1.7948$, a limit of
$N({\HI})<5 \times 10^{18}$ was inferred from the lack of damping
wings in an {\it IUE} spectrum (Bechtold \etal 1987\nocite{Bechtold87}).
Thus, $50$~\% of the HIRES/Keck very strong {\MgII} absorption systems
have known {\HI} column densities and both of them are not DLAs.

Rao \& Turnshek (2000\nocite{Rao00}) searched for DLAs in the {\it
HST}/FOS spectra of $87$ {\MgII}--absorption systems.  Of the $87$
{\MgII} absorbers in their sample, nine of them have $W_r >
1.8$~{\AA}.  Of those nine absorbers, four are at $z > 1$.  In total,
$4/5$ ($80$~\%) of the $z<1$ absorbers are DLAs, and $1/4$ ($25$~\%)
of the $z>1$ absorbers are DLAs.  This suggests not only that the
$z>1$ very strong {\MgII} systems are a different population than the
$z<1$ systems, but that the $z>1$ population displays a relative
paucity of DLAs.  Furthermore, in contrast to both superwinds and very
strong {\MgII} absorption systems, DLAs display no significant
evolution in $N(z)$ from $z=4$ to $z=0$ (\cite{Rao00};
\cite{cwc01}).  If, as we hypothesize, superwinds give rise to the
majority of $z > 1$, $W_r > 1.8$~{\AA} {\MgII} absorbers, we are
compelled to conclude that most DLAs are not superwinds.

\begin{figure*}
\figurenum{3}
\plotfiddle{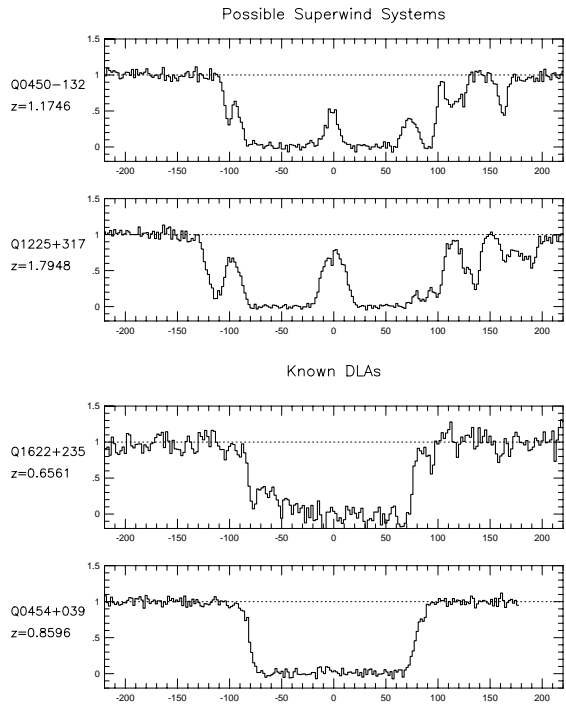}{6.in}{0}{40}{40}{-150}{-20}
\caption{
The top two panels show normalized {\MgII}~2796 absorption profiles
for two characteristic $W_r>1.8$~{\AA} profiles at $z>1$.  The
profiles show a central inversion, as would be expected if the line of
sight were passing through a superwind.  The lower two panels show
very strong {\MgII} absorbers at $z<1$ that are also damped {\Lya}
absorbers.  The DLA systems show a single saturated {\MgII}
absorption region and no central inversion.  }
\label{fig:DLA}
\end{figure*}

Why might this be the case?  Although large amounts of material ($\sim
60$~{\msun}~yr$^{-1}$) are expelled in a superwind, the majority of
this material is highly ionized (e.g. {\cite{Tomisaka93}},
{\cite{Suchkov94}}, {\cite{Strickland98}}, {\cite{TT98}}).  The only
clouds that gives rise to low--ionization absorption are the dense
clouds of entrained material.  High--equivalent width systems like the
ones presented in Figure~{\ref{fig:strong}} could be produced with
only small numbers of moderate--column density clouds (greater than
about five) because the clouds in a superwind have a significant
spread in velocity space ($\sim 300$~{\kms}).  It has been shown that
the majority of DLAs have total velocity spreads less than
$200$~{\kms} (\cite{Prochaska99}; \cite{Pettini00}), indicating that
they are typically produced in environments less active than those
which produce the very strong {\MgII} absorbers.  Figure~\ref{fig:DLA}
illustrates these differences.  The bottom two panels contain
high--resolution {\MgII} profiles of two DLAs (\cite{archive2}) and
the top two panels show examples of symmetric--inverted {\MgII}
profiles from our sample.

\section{Conclusion}
\label{sec:conclude}

The majority of very strong ($W_r>1.8$~{\AA}), $z>1$ {\MgII}
absorption profiles with high--resolution spectra show distinctive
kinematic features, including a central inversion and a large velocity
spread.  All of these features can be explained with a line of sight
through a superwind.  The redshift number density of absorbers
expected from superwinds, $N(z)$, was calculated using parameters
observed in local superwinds and was found to be consistent with the
$N(z)$ of very strong {\MgII} absorbers.

In addition, the absence of very strong {\MgII} absorbers is
consistent with the decrease in star formation from $z=1$ to the
present.  The number of DLAs, however, is consistent with no evolution
over this redshift range.  In addition, the very strong {\MgII}
absorbers at $z>1$ with known {\HI} column densities display a paucity
of DLAs as compared with the low--redshift very strong {\MgII}
absorbers, suggesting that the majority of DLAs do not arise in
superwinds.



\include{tab1}
\include{tab2}


\newpage

\newpage

\newpage
\end{document}

%% file: tab1.tex
\begin{deluxetable}{rcccrccc}
\tablewidth{0pc}
\tablecaption{Journal of Observations}
\tablehead
{
\colhead{Object} &
\colhead{V [mag]} &
\colhead{$z_{\rm em}$} &
\colhead{Date [UT]\tablenotemark{a}} &
\colhead{Exp [s]\tablenotemark{b}} &
\colhead{$\lambda $ Range [\AA]} &
\colhead{Decker} &
\colhead{Blocker}
}
\startdata
Q~$0450-132$ & 17.5 & 2.253 & 1995 Jan 24 &  5400 & 3986.5--6424.5 & 7{\arcsec} 
& kv370 \nl
Q~$1213-003$ & 17.0 & 2.691 & 1995 Jan 24 &  5200 & 5008.1--7356.7 &
14{\arcsec} & kv418 \nl
B2~$1225+317$ & 15.9 & 2.219 & 1995 Jan 24 &  2400 & 5737.5--8194.7 &
14{\arcsec} & og530 \nl
\enddata
\label{tab:obsjournal}
\end{deluxetable}

%% file: tab2.tex
\begin{deluxetable}{lcc}
\tablewidth{0pc}
\tablecaption{Absorber Properties}
\tablehead
{
\colhead{Ion/Tran} &
\colhead{$W_{r}$, {\AA}} &
\colhead{$N_{a}$, {\cmsq}} 
}
\startdata
\cutinhead{Q~$0450-132~~z_{\rm abs}=1.1746~~\Delta v = 296$~{\kms}}
 {\MnII}~2577  &  $ 0.14\pm0.05$ &  $ 12.87\pm 0.02$  \nl
 {\MnII}~2594  &  $ 0.09\pm0.04$ & \nl
 {\MnII}~2606  &  $ 0.06\pm0.04$ & \nl
 {\FeII}~2261  &  $ 0.13\pm0.04$ &  $ 14.99\pm 0.03$ \nl
 {\FeII}~2344  &  $ 0.77\pm0.02$ & \nl
 {\FeII}~2374  &  $ 0.49\pm0.03$ & \nl
 {\FeII}~2383  &  $ 1.07\pm0.02$ & \nl
 {\FeII}~2600  &  $ 1.15\pm0.02$ & \nl
 {\MgII}~2796  &  $ 1.82\pm0.02$ &  $>14.07         $  \nl
\cutinhead{Q~$1213-003~~z_{\rm abs}=1.3201~~\Delta v = 397$~{\kms}}
 {\MnII}~2577  &  $<0.08        $ &  $<11.77         $  \nl
 {\MnII}~2606  &  $<0.07        $ & \nl
 {\FeII}~2344  &  $ 1.07\pm0.09$ &  $ 14.53\pm 0.02$  \nl 
 {\FeII}~2374  &  $ 0.42\pm0.07$ & \nl
 {\FeII}~2383  &  $ 1.67\pm0.06$ & \nl
 {\FeII}~2587  &  $ 0.88\pm0.07$ & \nl
 {\FeII}~2600  &  $ 1.59\pm0.03$ & \nl
 {\MgII}~2796  &  $ 2.79\pm0.02$ &  $>14.53         $  \nl
 {\MgII}~2803  &  $ 2.51\pm0.02$ &  \nl
 {\MgI}~2853   &  $ 0.69\pm0.05$ &  $ 12.84\pm 0.01$  \nl
\cutinhead{Q~$1213-003~~z_{\rm abs}=1.5541~~\Delta v = 430$~{\kms}}
 {\MnII}~2594  &  $<0.06        $ &  $<12.18         $ \nl
 {\MnII}~2606  &  $<0.09        $ & \nl
 {\FeII}~2261  &  $ 0.02\pm0.05$ &  $ 14.40\pm 0.01$ \nl
 {\FeII}~2344  &  $ 0.50\pm0.06$ & \nl
 {\FeII}~2374  &  $ 0.28\pm0.05$ & \nl
 {\FeII}~2383  &  $ 0.85\pm0.05$ & \nl
 {\FeII}~2587  &  $ 0.46\pm0.06$ & \nl
 {\FeII}~2600  &  $ 0.85\pm0.05$ & \nl
 {\MgII}~2796  &  $ 1.99\pm0.03$ &  $ >14.19        $ \nl
 {\MgII}~2803  &  $ 1.48\pm0.04$ & \nl
 {\MgI}~2853   &  $ 0.22\pm0.08$ &  $ 12.31\pm 0.03$ \nl
\cutinhead{B2~$1225+317~~z_{\rm abs}=1.7948~~\Delta v = 343$~{\kms}}
 {\MnII}~2577  &  $<0.04$         &  $ 12.13\pm 0.14$    \nl
 {\MnII}~2606  &  $<0.05$         & \nl
 {\FeII}~2344  &  $ 0.30\pm0.04$ &  $ 13.88\pm 0.01$  \nl
 {\FeII}~2374  &  $ 0.09\pm0.04$ & \nl
 {\FeII}~2383  &  $ 0.66\pm0.03$ & \nl
 {\FeII}~2600  &  $ 0.62\pm0.05$ & \nl
 {\MgII}~2796  &  $ 2.01\pm0.02$ &  $ >14.29        $ \nl
 {\MgII}~2803  &  $ 1.59\pm0.03$ & \nl
 {\MgI}~2853   &  $ 0.28\pm0.05$ &  $ 12.40\pm 0.02$ \nl
\enddata
\label{tab:properties}
\end{deluxetable}